\documentclass[conference]{IEEEtran}

\usepackage{cite}      

\usepackage{graphicx}

\usepackage{url}       

\usepackage{amsmath}   
\makeatletter
\let\NAT@parse\undefined
\makeatother
\usepackage[hypertex]{hyperref}
\newtheorem{theorem}{Theorem}
\newtheorem{lemma}{Lemma}
\newtheorem{definition}{Definition}
\newcommand{\ulp}{\text{ulp}}
\usepackage[latin9]{inputenc} \usepackage[T1]{fontenc}
\usepackage{amsfonts}\newcommand{\R}{\mathbb{R}}\renewcommand{\Pr}{\mathbb{P}}\def\Ex{\mathbb{E}}
\usepackage{amssymb}
\usepackage{listings}
\begin{document}

\title{Stochastic Formal Methods: \\ An application to accuracy of numeric software}

\author{\authorblockN{Marc Daumas}
\authorblockA{%
  CNRS-LIRMM visiting LP2A\\
  University of Perpignan Via Domitia\\
  Perpignan, France 66860\\
  Email: Marc.Daumas@Univ-Perp.Fr}
\and
\authorblockN{David Lester}
\authorblockA{%
  School of Computer Science\\
  University of Manchester\\
  Manchester, United Kingdom M13 9PL\\
  Email: David.R.Lester@Manchester.Ac.UK}
}

\maketitle

\begin{abstract}
  This paper provides a bound on the number of numeric operations
  (fixed or floating point) that can safely be performed before
  accuracy is lost.  This work has important implications for control
  systems with safety-critical software, as these systems are now
  running fast enough and long enough for their errors to impact on
  their functionality. Furthermore, worst-case analysis would blindly
  advise the replacement of existing systems that have been
  successfully running for years.  We present here a set of formal
  theorems validated by the PVS proof assistant.  These theorems will
  allow code analyzing tools to produce formal certificates of
  accurate behavior.  For example, FAA regulations for aircraft
  require that the probability of an error be below $10^{-9}$ for a 10
  hour flight \cite{JohBut92}.
\end{abstract}

\IEEEpeerreviewmaketitle

\section{Introduction}

Formal proof assistants are used in areas where errors can cause
loss of life or significant financial damage as well as in areas where common
misunderstandings can falsify key assumptions. For this reason, formal
proof assistants have been much used in floating point arithmetic
\cite{Rus98,Har2Ka,BolDau03,DauMelMun05,MunLes05}. Previous references
just link to a few projects using proof assistants such as ACL2, HOL
\cite{GorMel93}, Coq \cite{HueKahPau04} and PVS \cite{OwrRusSha92}.

All these projects deal with worst case behavior. Recent work has
shown that worst case analysis is meaningless for applications that
run for a long time. For example, a process adds numbers in $\pm 1$ to
single precision, and therefore has a round-off error of $\pm
2^{-25}$.  If this process adds $2^{25}$ items, then the accumulated
error is $\pm 1$, and note that 10 hours of flight time at operating
frequency of 1~kHz is approximately $2^{25}$ operations.  Yet we
easily agree that provided the round-off errors are not correlated,
the actual accumulated error will be much smaller.

Developments in probability share many features with developments in
floating point arithmetic:
\begin{enumerate}
\item Each result usually relies on a long list of hypotheses. No
  hypothesis can be removed, but slight variations induce a large
  number of results that look almost identical.
\item Most people that use the results are not specialists in the
  specific field.  They want a trustworthy result but they are not
  proficient enough to either select the best scheme or detect minor
  faults that can quickly lead to huge problems.
\end{enumerate}

For these reasons, we are strongly of the opinion that validation of a
safety-critical numeric software using probability should be done
using an automatic proof checker. We present in
Section~\ref{sec/stoch} the model that we are using. Section
\ref{sec/prob} presents our formal developments in probability. The
Doobs-Kolmogorov inequality provides an effective way to compute the
probability that a piece of software will successfully run within an
acceptable error bound.

This work is connected to continuous space Markov random walks or
renewal-reward processes though these applications focus on asymptotic
behavior \cite{Bre98,Fuh04}. We want to precisely bound the
probability of remaining within bounds for a given number of steps.
This is connected to ruin probabilities \cite{Asm2K} and the
Doobs-Kolmogorov inequality for martingales \cite{GriSti82}. Related
work on theoretic construction of the probability space using higher
order logic can be found in \cite{Hur02,AudPau06} and references herein. In the
rest of this text, we assume that the created round-off and measure
errors are unbiased independent random variables or that their
expectation conditional to the previous errors is zero.

\section{Stochastic model}
\label{sec/stoch}

\subsection{Individual round-off errors of fixed and floating point
  operations}

We are dealing with fixed or floating point numbers. A floating point
number represents $v = m \times 2^e$ where $e$ is an integer and $m$
is a fixed point number \cite{Gol91}.  IEEE 754 standard \cite{Ste.87}
uses sign-magnitude notation for the mantissa and the first bit of the
mantissa is implicit in most cases leading to the following definition
where $s$ and all the $b_i$ are either 0 or 1 (bits).
$$
v = (-1)^s \times 1.b_1 \cdots b_{p-1} \times 2^e
$$
Some circuits such as TMS320 uses two's complement notation for $m$
leading to the following definition \cite{Tex97}.
$$
v = (1.b_1 \cdots b_{p-1} - 2 \times s) \times 2^e
$$

For both notations, we define for any representable number $x$, the
unit in the last place function where $e$ is the exponent of $x$ as
above. In fixed point notation, $e$ is a constant provided by the data
type.
$$
\ulp (v) = 2^{e - p + 1}
$$

A variable $v$ is set either by an external sensor or by an operation.
Trailing digits of numbers randomly chosen from a logarithmic
distribution \cite[p.  254-264]{Knu97} are approximately uniformly
distributed \cite{FelGoo76}. So we can assume that if $v$ is a data
obtained by an accurate sensor, the difference between $v$ and the
actual value $\overline{v}$ is uniformly distributed in the range $\pm
\ulp(v)/2$. We can model the representation error $v - \overline{v}$
by a random variable $X$ with expectation $\Ex(X) = 0$ and variance
$\Ex(X^2) = \ulp(v)^2/12$.  The sensor may be less accurate leading to
a larger variance but it should not be biased.

Round-off errors created by operators are discrete and they are not
necessarily distributed uniformly \cite{BusFelGooLin79}.  For each
operator $\circledast$ implementing the real operation $\ast$, we
define
$$X = V \circledast W - V \ast W$$
where $V$ and $W$ are number distributed logarithmically over
specified ranges. The distribution of $X$ is very specific but we
verify that the expectation is $\Ex(X) = 0$ and we bound its variance
$\Ex(X^2)$.

Fixed point additions do not create any additional round-off error
provided its output is in the same format as its inputs.  Reducing the
format of a fixed point number creates a uniformly distributed round
off error provided the input was logarithmically distributed
\cite{FelGoo76}.

\subsection{Round off errors of an accumulation loop}
\label{sub/loop}

We will use two examples. The first one is given in
listing~\ref{lst/int}. It sums data produced by a fixed point sensor 
$x_i$ with a measure error $X_i$.

\begin{lstlisting}[caption={Simple discrete integration from \cite{BriDauLanMar06}},firstnumber=1,label=lst/int]
$a_0 = 0$
for ($i = 0$; $i < n$; $i = i + 1$)
  $a_{i+1} = a_i + x_i$
\end{lstlisting}

We can safely assume that $X_i$ are independent identical uniformly
distributed random variables over $\pm \ulp(x_i)/2$. Data are fixed
point meaning that the sum $a_i + x_i$ does not introduce any rounding
error and the weigth of one unit in the last place does not depend on
$x_i$ so we write $\ulp$ instead of $\ulp(x_i)$.  After $n$
iterations, we want the probability that the accumulated measure error
have always been constrained into user specified bounds $\epsilon$.
Using the Doobs-Kolmogorov inequality of Theorem~\ref{theo/doobs}
where $S_i = \sum_{j=1}^i X_j$, we have that
$$\Pr\left(\max_{1\le i\le n}(|S_i|)\le\epsilon\right) \le 1 - \frac{n \ulp^2}{12 {\epsilon}^2}.$$

The second example is given in listing~\ref{lst/ode}. It solves
initial value problem (IVP) ordinary differential equations (ODE) by
computing an incremental slope $\Phi(t_i, h_i, x_i, f)$ based on the
current time $t_i$, the current step size $h_i$, the current value of
the function $x_i$ and the differential equation $x'(t) = f(t, x(t))$.
The function $\Phi$ may be very simple using Euler's explicit method
or more complex using any Runge-Kutta method or any implicit method.
We focus here on scalar ODEs although our analysis may apply to
vectors.  Line 4 assume for the sake of simplicity that $h_i$ is a
constant although this is not neccessary.

\begin{lstlisting}[caption={Solving initial value problem ordinary differential equations \cite{StoBur02}},firstnumber=1,label=lst/ode]
for ($i = 0$; $i < n$; $i = i + 1$) {
  $x_{i+1} = x_i + h_i \times \Phi(t_i, x_i, h_i, f)$
  $t_{i+1} = t_i + h_i$
  $h_{i+1} = h_i$
}
\end{lstlisting}

Our first guess was to introduce a sequence of random variables
$\{X_n\}$ that models the difference introduced by round-off errors at
step $i$. In most cases, $\Phi$ introduces a drift due to higher order
effect of random variables and a drifted correlation between the error
introduced at step $i+1$ and errors on the previous steps.  For
example, the square of a rounded value $v + V$ where $v$ is the stored
value and $V$ is a random variable, introduces a positive drift due to
$V^2$ term that is always positive. So we model the effect of the
round-off error by two terms $X_i$ and $Y_i$. We use the
Doobs-Kolomogorov inequality of Theorem~\ref{theo/doobs} for the
sequence $\{X_n\}$ and worst case error analysis for the sequence
$\{Y_n\}$ setting the following conditional expectation
$$
\Ex(X_n; X_1 \cdots X_{n-1}) = 0.
$$

Random variables $X_{i+1}$ and $Y_{i+1}$ account for the round-off and
propagated errors introduced by replacing
$$
x_i + X_i + Y_i + h_i \times \Phi(t_i, x_i + X_i + Y_i, h_i, f)
$$
with 
$$
x_i \oplus h_i \otimes \tilde{\Phi}(t_i, x_i, h_i, f)
$$
where $\tilde{\Phi}$ is evalaution of $\Phi$ in computer.  First order
effect of round-off errors created are accounted in $X_{i+1}$. Higher
order effect of round-off errors created and propagated effect of
$X_i$ and $Y_i$ in $\Phi$ are accounted in $Y_{i+1}$.

$\{X_n\}$ is constructed to contain only independent random variables with no drift
$\Ex(X_i) = 0$ and we only need to bound their variance $\Ex(X_i^2)$.
We will do worst case analysis on $\{Y_n\}$ and we bound each $Y_i$
with interval arithmetic \cite{JauKieDidWal01}. Software such as
Fluctuat \cite{GouMarPut06} is already able to distinguish between first order and higher
order error terms.

\section{Probability distribution of being safe}
\label{sec/prob}

\subsection{Probability}

We have two main choices in presenting an account of probability: one
is to take an informal approach, the second involves taking
foundational matters seriously. In this paper we will consistently try
to present matters informally except for Section~\ref{sub/form},
however the reader should be aware that the PVS system underlying
these results is built on the firm foundations for probability theory
(using measure theory) \cite{Hal44,Hal50}. A middle way between
extreme formality and an accessible level of informality is to be
found in \cite{GriSti82}.

We begin by defining the {\em distribution function} of a random
variable.
\begin{definition}
A random variable $X$ has {\em distribution function} $F$, if
$\Pr(X\leq x) = F(x)$
\end{definition}
As we will be studying {\em continuous random variables}, these are
defined as follows:
\begin{definition}
A random variable $X$ is {\em continuous} if its distribution function
can be expressed as
\[F(x) = \int_{-\infty}^x f(x) dx\]
for some integrable function $f:\R \to [0,\infty)$. We call the
function $f$ the probability density function for the random variable
$X$.
\end{definition}
As an example of a continuous random variable, consider the
temperature $T$ at a certain point in an industrial process. Even if
an attempt is being made to hold this temperature constant, there will
be minor fluctuations, and these can be modeled mathematically by
assuming that $T$ is a continuous random variable.

The other concept we will need is that of dependent and independent
random variables. Suppose we model the outcomes of the tossing of two
coins $C_1$ and $C_2$ by random variables. Provided there is nothing
underhand going on, we would expect the result of tossing the first
coin to have no effect on the result of the second coin, and {\it vice
versa}. If this is the case, then we say that $C_1$ and $C_2$ are {\em
independent}. Consider an alternative scenario in which having tossed
the coin $C_1$ and discovered that it has come up ``heads'', and we
now define the random variable $C_2$ to be: the outcome: ``the
downward facing side of the coin $C_1$ is tails''. In this case the
random variables $C_1$ and $C_2$ are {\em dependent}.

The other idea we must address is that of {\em conditional probability}.
\begin{definition}
We define the probability of ``$A$ given $B$'' (written $\Pr(A;~B)$) as:
\[\Pr(A;~B) = \frac{\Pr(A\mathrel{\cap}B)}{\Pr(B)}\]
whenever $\Pr(B)>0$.
\end{definition}
As an example: if event $A$ is ``I am carrying an umbrella'' and event
$B$ is ``it is raining'', then $Pr(A;~B)$ is the probability that ``I
am carrying an umbrella given that it is raining''. Note that although
in general $\Pr(A;~B) \not= \Pr(B;~A)$, in this particular case, if
you live in Perpignan or Manchester, then on most days: $\Pr(A;~B) =
\Pr(B;~A)$, though for rather different reasons.

\subsection{A Formal Development of probability}
\label{sub/form}

\begin{definition}
A {\em $\sigma$-algebra} over a type $T$, is a subset of the power-set
of $T$, which includes the empty set $\{\}$, and is closed under the
operations of complement, countable union and countable intersection.
\end{definition}

If $T$ is countable -- as it is for discrete random variables -- then
we may take $\sigma=\wp(T)$; if the set $T$ is the reals -- as it
is for continuous random variables -- then we make $\sigma=B$: the
Borel sets.

\begin{definition}
A {\em Measurable Space} $(T,\sigma)$ is a set (or in PVS a type) T,
and a {\em $\sigma$-algebra} over $T$.
\end{definition}

\begin{definition}
A function $\mu:\sigma\to{\R}_{\ge 0}$ is a {\em Measure} over the
$\sigma$-algebra $\sigma$, when $\mu(\{\}) = 0$, and for a sequence of disjoint
elements $\{E_n\}$ of $\sigma$:
\[\mu\left(\bigcup_{n=0}^\infty E_n\right) = \sum_{n=0}^\infty \mu(E_n).\]
\end{definition}

\begin{definition}
A {\em Measure Space} $(T,\sigma,\mu)$ is a measurable space
$(T,\sigma)$ equipped with a measure $\mu$.
\end{definition}

\begin{definition}
A {\em Probability Space} $(T,\sigma,\Pr)$ is a measure space
$(T,\sigma,\Pr)$ in which the measure $\Pr$ is finite for any set in
$\sigma$, and in which:
\[\Pr(X^c) = 1-\Pr(X).\]
\end{definition}

The PVS development of probability spaces in Figure~\ref{lst/prob},
takes three parameters: $T$, the sample space, $S$, a $\sigma$-algebra
of permitted events, and, $\Pr$, a probability measure, which assigns
to each permitted event in $S$, a probability between $0$ and $1$.
Properties of probability that are independent of the particular
details of $T$, $S$ and $\Pr$ are then provided in this file. If we
wished to discuss continuous random variables then we would partially
instantiate this PVS file with {\tt T = real}, and {\tt S =
  borel\_set}. If we go further and also specify $\Pr$, we will have
described the random variable distributions as well.  Of particular
interest later is the fact that the sum of two random variables is
itself a random variable, and consequently any finite sum of random
variables will be a random variable.

\begin{figure*}
  \begin{center}\fbox{\begin{minipage}{0.98\linewidth}
  {\footnotesize \input{proba_space}}
  \end{minipage}}\end{center}
  \caption{Abbreviated probability space file in PVS}
  \label{lst/prob}
\end{figure*}

\begin{definition}
If $(T_1,{\sigma}_1,{\Pr}_1)$ and $(T_2,{\sigma}_2,{\Pr}_2)$ are
probability spaces then we can construct a {\em product probability
space} $(T_3,{\sigma}_3,{\Pr}_3)$, where:
\[\begin{array}{lll}
T_3 &=& T_1 \times T_2 \\
{\sigma}_3 &=& \sigma({\sigma}_1\times {\sigma}_2)\\
{\Pr}_3'(a,b) &=& {\Pr}_1(a){\Pr}_2(b)
\end{array}\]
where ${\Pr}_3$ is the extension of ${\Pr}'_3$ that has the whole of
${\sigma}_3$ as its domain.
\end{definition}

Note that the product probability ${\Pr}_3$ has the effect of
declaring that the experiments carried out in probability spaces
$(T_1,{\sigma}_1,{\Pr}_1)$ and $(T_2,{\sigma}_2,{\Pr}_2)$ are
independent. Obviously, the process of forming products can be
extended to any finite product of finitely many probability
spaces. Currently, it is not clear whether PVS is powerful enough to
capture the notion of a countably infinite sequence of random
variables $\{X_n\}_{n=1}^{\infty}$; fortunately, in this work we don't
currently require this result.

In Figure~\ref{lst/cond}, we define the conditional probability
$\Pr(A;~B)$ (written {\tt P(A,B)} as PVS will not permit the use of
``;'' as an operator). We take the opportunity to prove Bayes' Theorem
along the way.

\begin{figure*}
  \begin{center}\fbox{\begin{minipage}{0.98\linewidth}
  {\footnotesize \input{conditional}}
  \end{minipage}}\end{center}
  \caption{Conditional probability file in PVS}
  \label{lst/cond}
\end{figure*}

\subsection{Continuous Uniform Random Variables}

If $X$ is a continuous random variable distributed uniformly over the
interval $[a,b]$, then informally it takes any value within the
interval $[a,b]$ with equal probability.

To make this more formal, we define the {\em characteristic function}
of a set $S$ as the function ${\chi}_S$, which takes the values $1$ or
$0$ depending on whether it is applied to a member of $S$.
\begin{definition}
  \[{\chi}_S(x) = \left\{\begin{array}{ll} 1 & x\in S\\
                                          0 & x\not\in S\end{array}\right.\]
\end{definition}
Now the probability density function $f$ of the uniform random
variable over the closed interval $[a,b]$ is
$\frac{1}{b-a}{\chi}_{(a,b]}$. From this we can calculate the
distribution function:
\[F(x) = \int_{-\infty}^x f(x) dx,\]
from which we can calculate the probability
\[\Pr(x<X<=y) = F(y)-F(x).\]

In the case where $X$ is distributed $U_{[0,1]}$, and because -- for
any $f(x)$ with $\int f = F$ -- we have
$$
\begin{array}{l}
  \displaystyle  \int_{-\infty}^\infty f(x) \chi_{(a,b]}(x) dx = \\
  \displaystyle  ~~~~~ (F(x)-F(a))\chi_{(a,b]}(x) + (F(b)-F(a))\chi_{(b,\infty)}(x).
\end{array}
$$

We also observe that if $X$ is distributed $U_{[a,b]}$, then $\Ex(X) =
\frac{a+b}{2}$, and $\mbox{Var}(X) = \frac{(a-b)^2}{12}$. So, with
$a=0$, $b=1$ we get: $\mu=\frac{1}{2}$, $\sigma^2 = \frac{1}{12}$.

\subsection{Sums of Continuous Random Variables}

\begin{definition}
If we have a sequence of continuous random variables $\{X_n\}$, then
we define their partial sums as a sequence of continuous random
variables $\{S_n\}$ with the property
\[S_n = \sum_{i=1}^n X_i.\]
\end{definition}

\begin{theorem}
If continuous random variables $X$ and $Y$ have joint probability density
functions $f$, then $Z=X+Y$ has probability
density function:
\[f_Z(z) = \int_{-\infty}^\infty f(x,z-x) dx.\]
\end{theorem}

In the special case where $X$ and $Y$ are independent, then (because
the joint probability density function $f(x,y)$ can be expressed as
the product $f_X(x)f_Y(y)$) we have the {\em Continuous Convolution
Theorem}:
\begin{theorem}
If continuous random variables $X$ and $Y$ are independent and have
probability density functions $f_X$ and $f_Y$ respectively, then
$Z=X+Y$ has probability density function:
\[f_Z(z) = \int_{-\infty}^\infty f_X(x)f_Y(z-x) dx
         = \int_{-\infty}^\infty f_X(z-x)f_Y(x) dx.\]
\end{theorem}

\subsection{Reliability of long calculations}

What we are actually interested in is whether a series of
calculations might accumulate a sufficiently large
error to become meaningless. In the language we have developed, we are
asking what is the probability that all calculations of
length upto $n$ is correct:
$$\Pr\left(\max_{1\le i\le n}(|S_i|)\le\epsilon\right).$$

Because they have nice convergence properties, we are especially
interested in {\em martingales}

\begin{definition}
  A sequence $\{S_n\}$ is a {\em martingale} with respect to the
  sequence $\{X_n\}$, if for all $n$:
  \begin{enumerate}
  \item $\Ex(|S_n|)<\infty$; and
  \item $\Ex(S_{n+1};~X_1,~X_2,~\ldots,~X_n) = S_n$
  \end{enumerate}
\end{definition}

We first observe that the sequence $S_n=\Sigma_{i=1}^n X_i$ (as
previously defined) is a martingale with respect to the sequence
$\{X_n\}$.
\begin{lemma}
  The sequence $\{S_n\}$, where $S_n=\sum_{i=1}^n X_i$, and each
  $X_n$ is an independent random variable with $\Ex(X_n)=0$, is
  martingale with respect to the sequence $\{X_n\}$.
\end{lemma}

Alternatively as could be needed for program \ref{lst/ode}:

\begin{lemma}
  The sequence $\{S_n\}$, where $S_n=\sum_{i=1}^n X_i$, and $\{X_n\}$
  satisfies for all $i$
  $$
  \begin{array}{r c l}
    \Ex(X_i) & = & 0 \\
    \Ex(X_i;~X_1 \cdots X_{i-1}) & = & 0,
  \end{array}
  $$
  the sequence $\{S_n\}$ is martingale with respect to the sequence
  $\{X_n\}$.
\end{lemma}

We now make use of the Doobs-Kolmogorov Inequality presented
Figure~\ref{lst/doobs}. The statement of Theorem~\ref{theo/doobs} is
deceptively simple. The key as the astute reader will observe is that
we have a restricted form of the Doobs-Kolmogorov Inequality in which
the sample spaces of the underlying sequence of random variables are
identical.  This is an artifact of the PVS type system which would
require us to prove multiple version of the theorem at each tuple of
instantiated types.

Although the type system used in PVS is extraordinarily flexible, it
is not as malleable as that used by professional mathematicians. To
capture mathematics in its entirety using a theorem prover, we would
need to dispense with any form of type checking\footnote{A weak form of
  type consistency is used in category theory, but this is so weak
  that we can introduce the Russel Paradox.}. For its intended use as
an aide to proving programs correct, this would fatally weaken PVS as
a useful tool. In addition, in many practice areas of mathematics, the
full generality of categorical constructs is an unnecessary luxury,
albeit one with a seductive, siren-like, appeal.

\begin{figure*}
  \begin{center}\fbox{\begin{minipage}{0.98\linewidth}
  {\footnotesize \input{doobs}}
  \end{minipage}}\end{center}
  \caption{Doobs-Kolmogorov inequality in PVS}
  \label{lst/doobs}
\end{figure*}

\begin{theorem}[Doobs-Kolmogorov Inequality]
  \label{theo/doobs}
  If $\{S_n\}$ is a martingale with respect to $\{X_n\}$ then,
  provided that $\epsilon>0$:
  \[\Pr\left(\max_{1\le i\le n}(|S_i|)\ge\epsilon\right)
         \le \frac{1}{{\epsilon}^2} \Ex(S_n^2)\]
\end{theorem}

In our particular case where each $X_i$ is an independent random
variable with $\Ex(X_i)=0$, and $\mbox{Var}(X_i)=\sigma_i^2$, we
observe that
$$
\Pr\left(\max_{1\le i\le n}(|S_i|)\le\epsilon\right)
 \ge 1-\frac{1}{{\epsilon}^2}\sum_{i=1}^n\sigma_i^2
$$

The short conclusion is therefore that eventually errors will
accumulate and overwhelm the accuracy of any numerical software.
However, if $\epsilon$ is large enough and each of the $\sigma_i^2$
are small enough, then the number of iterations required for this to
occur will be high enough to be of no practical significance.
Crucially, the results hinge critically on the errors $\{X_n\}$ being
independent.

\section{Future work}

This work will be continued in two directions. The first direction is
to modify Fluctuat to generate theorems that can be checked
automatically by PVS using
ProofLite\footnote{\url{http://research.nianet.org/~munoz/ProofLite/}.}
as proposed in \cite{DauMelMun05,MunLes05}. This work will be carried
in collaboration with the developers of Fluctuat. The software will
conservatively estimate the final effect of the error introduces by
each individual floating point operations and compute upper bounds of
their variances.

The second direction is to develop and check accurate proofs about the
round-off errors of individual equations.  A uniformly distributed
random variable whose variance depends only on the operation and the
computed result might provide a too pessimistic bound. For example the
floating point addition of a large number with a small number absorbs
the small number meaning that the round-off error may be far below
half an ulp of the computed result.

Two's complement operation of TMS320 circuit can either round or
truncate the result. If truncation is used, it introduces a drift and
Doobs-Kolmogorov inequality for martingales cannot be used. Should we
wish to extend this work to account for drifts (non-zero means for the
random variables $\{X_n\}$), then we anticipate making use of Wald
Identity. Such developments will also be necessary to address higher
order error terms that introduce a drift.

This library and future work will be included into NASA Langley PVS
library\footnote{\url{http://shemesh.larc.nasa.gov/fm/ftp/larc/PVS-library/pvslib.html}.}
as soon as it becomes stable.

We saw with the example of listing~\ref{lst/ode} that inductions on
the variances of the random variables can be crudely bounded. Yet, we
may expect tighter results if we use tools that are able to infer
inductions and solve them mathematically but this domain is far from
the authors' research areas.

\section{Conclusions}

To the best of our knowledge this paper presents the first application
of the Doobs-Kolmogorov Inequality to software reliability and the
first generic formal development able to handle continuous, discrete
and non-continuous non-discrete random variable for PVS. Previous
developments in higher order logic where targeting other applications
and using Coq, HOL or Mizar proof assistants (see \cite{Hur02,AudPau06} and
references herein).  In addition, we have demonstrated a slightly
weaker version of this result in PVS. We claim that the utility of
this weaker result is not unduly restrictive, when compared to the
general result. The major restriction lies in the fact that we have to
demonstrate that a sequence of overall errors is martingale with
respect to the sequence of individual errors. We have been forced to
make simplifications to the mathematical model of our software to
ensure that this is the case. In particular, we have been forced to
insist that our individual errors have no drift, and are independent.

We have been surprised that the limit on the reliability of a piece of
numeric software could be expressed so succinctly. Notice that even
with a high tolerance of error, and with independent errors, we will
still eventually fail. Our results permit the development of safe upper
limits on the number of operations that a piece of numeric software
should be permitted to undertake.

It is worth pointing out that violating our assumptions (independence
of errors, and zero drift) would lead to worse results, so one should
treat the limits we have deduced with caution, should these
assumptions not be met.

\section*{Acknowledgment}

This work has been partially funded by CNRS PICS 2533 and was
partially done while one of the authors was an invited professor at
the University of Perpignan Via Domitia. It benefits from links
between the École Normale Supérieure de Lyon where one
author used to work and the University of Manchester started in the
Mathlogaps multi-participant Early Stage Research Training network of
the European Union. The authors would like to thanks Philippe Langlois,
Harold Simmons and Jean-Marc Vincent for fruitful informal discussions
on this work.

\bibliographystyle{IEEEtran.bst}
\bibliography{alias,perso,groupe,saao,these,livre,arith}

\end{document}